К.В. Моисеев
НИУ «Высшая школа экономики», Нижний Новгород

# Моделирование перехода России от солидарной пенсионной системы к накопительной[1]

**Аннотация.** В странах с растущим числом пожилых и сокращающейся рабочей силой, одной из которых является и Россия, становится невозможным поддерживать солидарную пенсионную систему, и возникает необходимость перейти к более устойчивой накопительной системе. В этой работе анализируются различные сценарии перехода России к такой системе. Это — первое исследование экономики России, в рамках которого строится OLG-модель для моделирования пенсионного перехода. Демонстрируется, что в долгосрочной перспективе переход к накопительной системе незначительно сокращает благосостояние пенсионеров. Однако на время перехода положение пенсионеров сильно ухудшается, в особенности при финансировании пенсий за счет налога на сбережения. Не менее важен и вывод о том, что переход накладывает тяжелое бремя на все поколения, живущие во время проведения реформы. Они вынужденно потребляют меньше и сильно изменяют схемы накопления и при этом также часто начинают больше работать. Подобные выводы делаются относительно среднестатистических когорт населения, результаты могут не совпадать для разных групп индивидов внутри когорт. В зависимости от механизма реформы пенсионный переход может вызвать как экономический рост, так и экономический спад, а также соответствующий им рост заработных плат и потребления или — их сокращение.

**Ключевые слова:** *пенсионная реформа, накопительная система, солидарная система, динамическая модель, OLG-model.*

Классификация JEL: D58, E2, E6.

### Введение

Переход от солидарной пенсионной системы к накопительной и его последствия стоят на повестке дня многих стран мира. Это связано с тем, что доминирующая сегодня по всему миру солидарная пенсионная система, основанная на государственных выплатах пенсионерам, изжила себя.

Солидарная система пенсионных выплат была создана в большинстве стран в начале прошлого века. Во время «бэби бума» она стала основной, и именно тогда сформировались ее главные черты. Выплаты пенсий нынешним пенсионерам в данной системе полностью зависят от собираемых с работающего населения налогов. Подобная организация системы пенсионных выплат сложилась в эпоху, когда на рынке труда было большое число работающего населения и при этом мало пенсионеров.

Сегодня же — все наоборот, коэффициент демографической нагрузки пожилого населения (отношение числа людей старше 65 к численности населения от 15 до 65) значительно выше, чем в середине прошлого века, а к 2030 г.

---

в РФ ожидается его увеличение до 50% и более — в соответствии с прогнозами Росстата[2]. В таких условиях невозможно поддерживать солидарную пенсионную систему в ее текущем виде, так как изымаемых у нынешнего сокращающегося работающего населения денег не хватит для поддержания благосостояния растущего числа пенсионеров. Необходимыми становятся либо сокращения выплат пенсионерам и повышение пенсионного возраста вместе со значительным увеличением налогов, либо переход к принципиально другой пенсионной системе.

Такой является частная накопительная система, в рамках которой не существует государственных пенсий и пенсионного налога, пожилые индивиды обеспечивают себя сами из своих сбережений, накопленных за всю их жизнь.

Таким образом, объектом исследования в данной работе являются основные факторы, изменяющиеся под воздействием перехода между государственной (солидарной) и частной (накопительной) пенсионной системами. В первую очередь речь идет о производстве, потреблении, сбережениях и предложении труда. В целях исследования для российской экономики будет построена модель общего равновесия пересекающихся поколений с эндогенным предложением труда, с совершенными ожиданиями и с государственным налоговым сектором. На ее основе будут рассмотрены изменения вышеназванных факторов в различных сценариях перехода от солидарной пенсионной системы к накопительной.

Также следует отметить, что Россия уже вставала на путь перехода к подобной системе в 2002 г. Введенная система, однако, сразу же оказалась несостоятельной и часто критиковалась за отсутствие контроля граждан над своими накоплениями, за чрезмерную сложность и непрозрачность для работников и плохое управление государственным пенсионным фондом ПФР (Арнаутова, Бекренёв, 2016). За более чем десятилетний период ее существования было предложено множество способов улучшить и расширить систему (Ahuja, Yermo, McTaggart, 2013). Однако в 2014 г. эта система, по сути, прекратила существование.

В 2014 г. был принят закон об отмене отчисления 6 из 22% страховых взносов с заработной платы каждого работника в накопительные пенсионные фонды. Официально такое решение было принято из-за неэффективности накопительных фондов. Так, в частности, во многих выступлениях региональных председателей ПФР заявлялось, что в 20 крупнейших негосударственных пенсионных фондах (НПФ), в которых сосредоточено более 70% пенсионных накоплений, средний показатель прироста вкладов составил от 2 до 8,3% в год при среднем показателе уровня инфляции за этот период — 9,65% в год[3]. Однако данная причина отмены пенсионных накоплений вызывает сомнения, так как в действительности пенсионные фонды могли быть и были доходными. Например, в (Мищенко, Леонидова, 2018) показано, что рост многих НПФ в период с 2009 по 2015 г. обгонял инфляцию, притом некоторых — значительно. Среди 11 крупнейших НПФ четыре обгоняли инфляцию, такое малое число доходных фондов среди крупнейших объясняется крайне высокой инфляцией за период до 2015 г. Например, в 2016 г., когда инфляция была достаточно низкой, доходность почти всех НПФ оказалась выше темпов роста потребительских цен. В целом, за период с 2009 по 2015 г. средняя доходность всех НПФ составила 76,6% при накопленной инфляции 79,1%. Несмотря на нестабильность цен в этот период, средняя доходность НПФ оказалась достаточно близка к темпам роста цен, что может свидетельствовать о том, что в условиях стабильной инфляции, как, например, в 2016 г., рынок консервативных финансовых инструментов способен обеспечивать доходность инфляции.

---

[2] Численность населения Российской Федерации по полу и возрасту. Росстат, 2022. Стат. бюллетень.

[3] https://sfr.gov.ru/branches/alania/news/interviev/~2014/08/19/404; https://sfr.gov.ru/branches/komi/news/interviev/~2014/08/08/126

Таким образом, российский опыт существования в парадигме частично накопительных пенсионных систем едва ли можно назвать серьезной попыткой перейти от солидарной системы к фондированной. А значит, и вопрос реалистичного и выполнимого полного перехода к накопительной системе все еще остается актуальным для Российской Федерации.

В зарубежной периодике существует немало подробных работ, рассматривающих математическое моделирование экономики во время пенсионного перехода при помощи модели пересекающихся поколений. В исследовании (Makarski, Hagemejer, Tyrowicz, 2017) поднимается вопрос о наиболее эффективной компенсации налогового разрыва после перехода к накопительной системе. В работах (Andersen, Bhattacharya, Liu, 2021; Andersen, Bhattacharya, 2017) рассматривается динамическая и Парето-эффективность солидарной и накопительной систем. Вопрос об эффективности на примере экономики США также поднимается в (Nishiyama, Smetters, 2007). В статье (Lin, Tanaka, Wu, 2021) наиболее полно моделируется и описывается динамика перехода к накопительной системе на примере экономики Тайваня. В работе (Song et al., 2015) детально анализируется перераспределение богатства между поколениями во времени.

В отечественной литературе данная тема представлена слабо, комплексные же исследования с применением моделирования отсутствуют.

В разд. 1 даются определения ключевым понятиям. В разд. 2 формулируется модель, дается определение функции полезности домохозяйств от потребления и труда, задается бюджетное ограничение. Затем рассматривается производственная функция Кобба–Дугласа, задача максимизации прибыли фирм. Далее формулируются условия адекватности в виде условий равенств на рынке спроса и предложения. В конце выводятся динамические уравнения Эйлера, модель полностью решается и находится равновесие; дается определение стационарному равновесному состоянию. В разд. 3 методом наименьших квадратов оцениваются параметры для производственной функции Кобба–Дугласа. Из литературы подбираются параметры для функции полезности от потребления; обобщенным методом моментов находятся параметры для функции полезности от труда; институциональные параметры задаются на основе теории, и этому выбору придается мотивация. Далее рассматривается качество подгонки параметров производственной функции. В разд. 4 сначала рассматривается качество подгонки модели в изначальном стационарном равновесии, затем анализируются изменения агрегированных показателей после реформы в сравнении с экономикой в изначальном равновесии. Далее в деталях объясняется динамика переходного пути ключевых показателей благосостояния: производства, сбережений, потребления и предложения труда. В разд. 5 подводятся итоги.

**1. Теоретические аспекты пенсионных систем**

Солидарная пенсионная система — вид государственной пенсионной системы, в которой определяется связь между поколениями. В такой системе государственные выплаты пенсионерам финансируются за счет отчислений с доходов нынешних работников.

Накопительная пенсионная система работает на принципе установленных пенсионных взносов, т.е. накопления, созданные взносами на протяжении жизни, выплачиваются пенсионеру после его выхода на пенсию. В накопительной пенсионной системе накопления аккумулируются на индивидуальном пенсионном счете, который находится в банках или пенсионных фондах.

Переход от солидарной пенсионной системы к накопительной подразумевает, что после него работники более не будут выплачивать часть своих доходов в государственный фонд для обеспечения пенсионеров, вместо этого каждый будет иметь индивидуальный накопительный счет в одном из банков, где на протяжении всего периода его занятости будут копиться сбережения, которые он будет регулярно отчислять из своих доходов.

Однако процесс перехода бывает разным. Это может быть как резкая отмена всяческих пенсий, так и постепенный переход на накопительную систему. Соответственно, макроэкономические показатели и показатели благосостояния в разных сценариях будут изменяться по-разному.

**2. Модель**

Используемая модель представляет собой модель пересекающихся поколений (OLG — overlapping generations), индивидов, живущих $S$ периодов, с совершенными ожиданиями, эндогенным предложением труда и государственным налоговым сектором. Она основана на модели из (Evans, DeBacker, 2017).

В модели нет роста населения, потому что в модели рассматриваются возрастные группы, в каждом периоде их число равно $S$, а уровень смертности индивида (или же возрастной группы, в данном случае это синонимы) равен 0 в каждом периоде, кроме последнего, где он равен 1.

**2.1.** Домохозяйства

Некоторое число идентичных индивидов рождается каждый период и живет $S$ периодов. Пускай возраст индивида будет обозначен как $s = \{1, ..., S\}$. Каждый индивид в каждый период $t$ наделен единицей времени $\tilde{l}$ и может потратить это время либо на труд $n_{s,t} \in [0, \tilde{l}]$, либо на отдых $l_{s,t} \in [0, \tilde{l}]$; при этом $l_{s,t} + n_{s,t} = \tilde{l}$ $\forall s,t$. Число когорт-поколений $S$ остается неизменным, в модели отсутствует демографическая компонента.

В каждом периоде своей жизни индивиды выбирают наборы потребления $\{c_{s,t+s-1}\}_{s=1}^{S}$, предложения труда $\{n_{s,t+s-1}\}_{s=1}^{S}$ и сбережений $\{b_{s+1,t+s}\}_{s=1}^{S-1}$ так, чтобы максимизировать свою пожизненную функцию полезности

$$\max_{\{c_{s,t+s-1},\, n_{s,t+s-1}\}_{s=1}^{S},\, \{b_{s+1,t+s}\}_{s=1}^{S-1}} \sum_{s=1}^{S} \beta^{s-1} u(c_{s,t+s-1},\, n_{s,t+s-1}), \qquad (1)$$

состоящую из попериодных функций полезности

$$u(c_{s,t}, n_{s,t}) = \frac{(c_{s,t})^{1-\sigma}}{1-\sigma} + \chi_s^n b \left[1 - \left(\frac{n_{s,t}}{\tilde{l}}\right)^\upsilon \right]^{1/\upsilon}, \qquad (2)$$

ограничивая себя попериодным бюджетным ограничением:

$$c_{s,t} + b_{s+1,t+1} = (1-\tau_{s,t}^l)w_t n_{s,t} + (1+r_t(1-\tau_{s,t}^k))b_{s,t} + X_{s,t}, \qquad (3)$$

где β — субъективный коэффициент дисконтирования; σ — коэффициент относительного неприятия риска (Wakker, 2008); $b$ и $\upsilon$ — параметры эллиптической полезности труда, аппроксимирующие уравнение (Evans, Phillips, 2017):

$$u(n_{s,t}) = -\chi_s^n (n_{s,t})^{1+1/\theta}/(1+1/\theta), \qquad (4)$$

$\chi_s^n$ — параметр, отвечающий за относительную важность полезности труда в сравнении с полезностью потребления; $b_{s+1,t+1}$ — сбережения индивида $s$ в периоде $t$, которые должны вернуться к нему с процентом в следующем периоде; $b_{s,t}$ — сбережения, с которыми индивид $s$ вошел в период $t$ и которые, соответственно, были выбраны в периоде $t-1$. Текущие заработные платы $w_t$ и ставка процента $r_t$ — одинаковы для всех индивидов. Доходы от труда и процентов по вкладам облагаются налогами по ставкам $\tau_{s,t}^l \geq 0$ и $\tau_{s,t}^k \geq 0$ соответственно, они могут варьировать в зависимости от периода $t$ и от когорты $s$. $X_{s,t}$ — трансферт индивиду $s$ в период $t$; с его помощью осуществляются выплаты пенсий в данной модели.

В силу предположения о сбалансированности государственного бюджета имеем $X_t = (S-R)^{-1} T_t$, где $R$ — пенсионный возраст, а $T_t$ — величина всех собранных за период $t$ налогов.

Второе слагаемое функции попериодной полезности (2), т.е. эллиптическая функция антиполезности труда (Evans, Phillips, 2017), с целью соблюдений условий Инады на всем множестве значений аппроксимирует функцию постоянной эластичности труда Фриша (constant Frisch elasticity, CFE) (Frisch, 1959) (см. (4)), где $\theta > 0$ — коэффициент эластичности предложения труда Фриша.

### 2.2. Фирмы

В экономике существует определенное число идентичных, совершенно конкурентных фирм, которые арендуют инвестиционный капитал $K_t$ у индивидов под реальный процент $r_t$ и нанимают работников за реальную заработную плату $w_t$. Фирмы используют их капитал $K_t$ и рабочую силу $L_t$ для выпуска продукции $Y_t$ в каждом периоде в соответствии с производственной функцией Кобба–Дугласа

$$Y_t = AK_t^\alpha L_t^{1-\alpha}, \quad \alpha \in (0,1), \quad A > 0, \qquad (5)$$

где α — доля капитала в доходе, а $A$ — общая производительность факторов производства.

Предполагается, что цена единицы выпуска в каждом периоде равна 1 ($P_t = 1$). Каждая фирма выбирает, какой объем капитала арендовать и как много работников нанять для максимизации функции налогооблагаемой прибыли

$$\max \pi = \max_{K_t, L_t}(1-\tau_t^c)(AK_t^\alpha L_t^{1-\alpha} - w_t L_t - \delta K_t) - r_t K_t, \qquad (6)$$

где $\delta \in [0,1]$ — коэффициент амортизации капитала; $\tau_t^c$ — ставка налога на бухгалтерскую прибыль, поэтому расходы на оплату труда и амортизационные отчисления подлежат вычету, а выплаты процентов по капиталу нет.

**2.3. Равенство спроса и предложения на рынках**

В этой модели на рынках труда, капитала и товара устанавливается равновесие спроса и предложения:

$$L_t = \sum_{s=1}^{S} n_{s,t} \quad \forall t, \tag{7}$$

$$K_t = \sum_{s=2}^{S} b_{s,t} \quad \forall t, \tag{8}$$

$$Y_t = C_t + I_t \quad \forall t, \tag{9}$$

$$I_t = K_{t+1} - (1-\delta)K_t.$$

А также соблюдается условие сбалансированности бюджета

$$X_t = \frac{1}{S-R}\sum_{s=1}^{S} \tau_{s,t}^l w_t n_{s,t} + \tau_{s,t}^k r_t b_{s,t} + \tau_t^c \left( Y_t - w_t L_t - \delta K_t \right), \tag{10}$$

где $R$ — пенсионный возраст.

**2.4. Стационарное равновесие системы уравнений**

Подстановка условий равенства спроса предложению (7) и (8) в уравнения максимизации прибыли фирм на основе (6) показывает, что в равновесии заработная плата и ставка процента — функции от распределения капитала $\Gamma_t = \{b_{s,t}\}_{s=2}^{S}$:

$$w_t(\Gamma_t) = (1-\alpha) A \left( \sum_{s=2}^{S} b_{s,t} \Big/ \sum_{s=1}^{S} n_{s,t} \right)^{\alpha}, \tag{11}$$

$$r_t(\Gamma_t) = (1-\tau_t^c)\left( \alpha A \left( \sum_{s=1}^{S} n_{s,t} \Big/ \sum_{s=2}^{S} b_{s,t} \right)^{1-\alpha} - \delta \right). \tag{12}$$

Максимизация пожизненной функции полезности (1) приводит к динамическим уравнениям Эйлера при $s \in \{1, ..., S\}$:

$$\frac{\partial u\left(\{b_{s+1,t+s}\}_{s=1}^{S-1}, \{n_{s,t+s-1}\}_{s=1}^{S}\right)}{\partial n_{s,t}} = 0 \Rightarrow w_t(1-\tau_{s,t}^l)(c_{s,t})^{-\sigma} = \frac{\chi_s^n b}{\tilde{l}}\left(\frac{n_{s,t}}{\tilde{l}}\right)^{\upsilon-1}\left[1-\left(\frac{n_{s,t}}{\tilde{l}}\right)^{\upsilon}\right]^{(1-\upsilon)/\upsilon}, \tag{13}$$

$$\frac{\partial u\left(\{b_{s+1,t+s}\}_{s=1}^{S-1}, \{n_{s,t+s-1}\}_{s=1}^{S}\right)}{\partial b_{s+1,t+1}} = 0 \Rightarrow (c_{s,t})^{-\sigma} = \beta\left(1 + r_{t+1}\left(1-\tau_{s,t+1}^k\right)\right)(c_{s+1,t+1})^{-\sigma}. \tag{14}$$

Подстановка (11), (12) и (3) в уравнения Эйлера (13) и (14) дает систему из $2S-1$ уравнений. Продленная на все периоды времени $t$ эта система полностью характеризует равновесие:

$$w_t(\Gamma_t)(1-\tau_{s,t}^l)\left(w_t(\Gamma_t)n_{s,t}(1-\tau_{s,t}^l) + \left[1 + r_t(\Gamma_t)(1-\tau_{s,t}^k)\right]b_{s,t} + X_t - b_{s+1,t+1}\right)^{-\sigma} =$$
$$= \frac{\chi_s^n b}{\tilde{l}}\left(\frac{n_{s,t}}{\tilde{l}}\right)^{\upsilon-1}\left[1-\left(\frac{n_{s,t}}{\tilde{l}}\right)^{\upsilon}\right]^{(1-\upsilon)/\upsilon}, \tag{15}$$

$$\left(w_t(\Gamma_t)n_{s,t}(1-\tau_{s,t}^l) + \left[1 + r_t(\Gamma_t)(1-\tau_{s,t}^k)\right]b_{s,t} + X_t - b_{s+1,t+1}\right)^{-\sigma} = \beta\left[1 + r_{t+1}(\Gamma_{t+1})(1-\tau_{s,t+1}^k)\right] \times$$
$$\times \left(w_{t+1}(\Gamma_{t+1})n_{s+1,t+1}(1-\tau_{s,t+1}^l) + \left[1 + r_{t+1}(\Gamma_{t+1})(1-\tau_{s,t+1}^k)\right]b_{s+1,t+1} + X_{t+1} - b_{s+2,t+2}\right)^{-\sigma}, \tag{16}$$
$$s \in \{1, ..., S-1\} \; \forall t.$$

Из этих уравнений можно найти равновесное стационарное состояние системы уравнений. Обозначим его условной переменной $x_t$ (т.е. $x_t = x_{t+1} = \bar{x}$).

**Определение 1** (равновесное стационарное состояние). Равновесное стационарное состояние в модели пересекающихся поколений индивидов, живущих $S$ периодов, с совершенными ожиданиями и эндогенным предложением труда, определено константным во всех периодах распределением потребления $\{\overline{c}_s\}_{s=1}^{s}$; индивидуальным предложением труда $\{\overline{n}_s\}_{s=1}^{s}$ и сбережений $\{\overline{b}_s\}_{s=2}^{s}$, а также заработной платой $\overline{w}$ и процентной ставкой $\overline{r}$ таким образом, что: 1) домохозяйства оптимизируют (13) и (14); 2) фирмы оптимизируют (6); 3) спрос и предложение уравновешиваются (7), (8).

### 3. Калибровка модели

**3.1.** Эконометрическая оценка параметров $\{A, \alpha\}$

В уравнении Кобба–Дугласа с постоянным масштабом (5) существует ограничение на параметр $0 < \alpha < 1$, так как $\alpha + (1-\alpha) = 1$. Для того чтобы корректно оценить параметры производственной функции (5), необходимо провести следующие преобразования:

$$Y_t = A K_t^\alpha L_t^{1-\alpha} \Rightarrow \ln Y_t = \ln A + \alpha \ln K_t + (1-\alpha) \ln L_t, \tag{17}$$

$$\ln Y_{t-1} = \ln A + \alpha \ln K_{t-1} + (1-\alpha) \ln L_{t-1}. \tag{18}$$

Вычтем из прологарифмированного уравнения (17) в периоде $t$ его версию в периоде $(t-1)$ (18):

$$\Delta \ln Y_t = \alpha \Delta \ln K_t + (1-\alpha) \Delta \ln L_t \Rightarrow [\Delta \ln Y_t - \Delta \ln L_t] = \alpha [\Delta \ln K_t - \Delta \ln L_t]. \tag{19}$$

Уравнение (19) можно оценить методом наименьших квадратов. Получив оценку параметра $\alpha + (1-\alpha) = 1$, можно найти и оценку параметра $A$.

$$\ln Y_t = \ln A + \alpha \ln K_t + (1-\alpha) \ln L_t \Rightarrow \ln Y_t - \alpha \ln K_t - (1-\alpha) \ln L_t = \ln(A).$$

Тогда

$$A = \exp(\ln Y_t - \alpha \ln K_t - (1-\alpha) \ln L_t). \tag{20}$$

Заменив константу $A$ из регрессии на константу (20), можно получить ее оценку.

Для построения регрессий (19) и (20) были выбраны следующие данные за период с 1999 по 2021 г.:

1) валовой внутренний продукт в ценах 1999 г. в качестве переменной агрегированного выпуска $Y_t$;

2) валовое накопление основного капитала в ценах 1999 г. в качестве переменной агрегированного капитала $K_t$;

3) численность занятых в качестве переменной агрегированного предложение труда $L_t$.

Как регрессор $[\Delta \ln K_t - \Delta \ln L_t]$, так и регрессант $[\Delta \ln Y_t - \Delta \ln L_t]$ из уравнения (19) стационарны в соответствии с тестом Дики–Фуллера, p-value-теста $1{,}46 \times 10^{-29}$ и $0{,}0012$ для регрессора и зависимой переменной соответственно.

Оценка регрессии (19) методом наименьших квадратов показала, что $\alpha = 0{,}3573$. Оценка $A$ из уравнения (20) — равносильна

$$A = \frac{1}{n} \sum_{t=1}^{N} \exp(\ln Y_t - \hat{\alpha} \ln K_t - (1-\hat{\alpha}) \ln L_t) \tag{21}$$

с подстановкой оценки $\alpha$. Тогда общая производительность факторов производства $A = 1{,}889$.

Оценки для $\alpha$ и для $A$ позволяют сформировать модель для предсказания $Y_t$ (рис. 1) и $r_t$ (рис. 2). Рассмотрим качество подгонки таких моделей с полученными ранее оценками $A = 1{,}889$ и $\alpha = 0{,}357$ для экономики России.

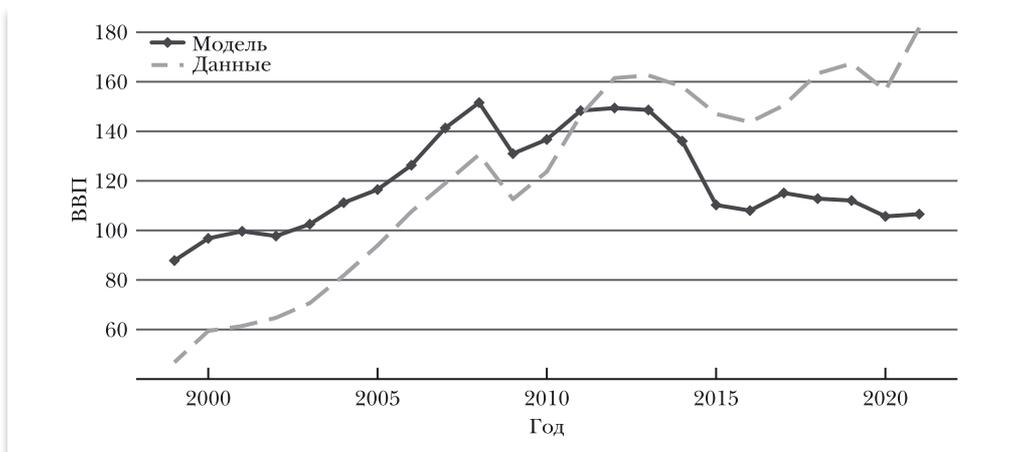

**Рис. 1.**
*Агрегированный выпуск РФ в ценах 1999 г.*

*Источник*: Росстат.

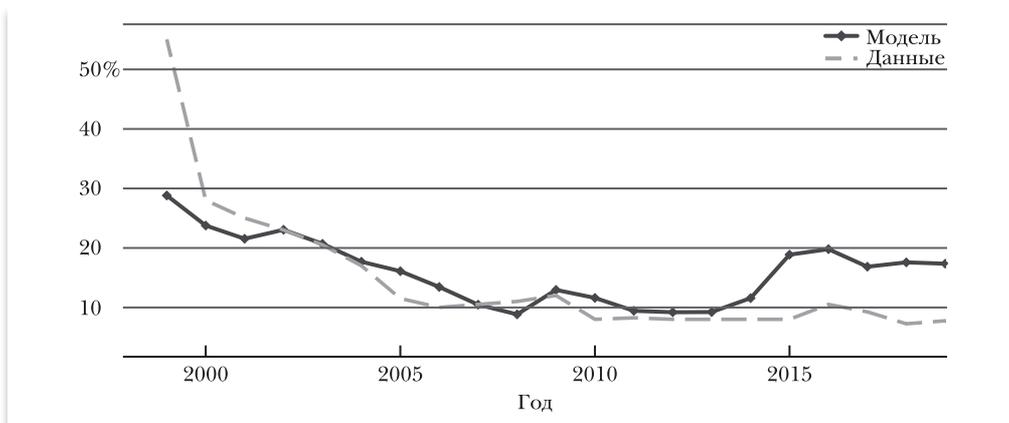

**Рис. 2.**
*Ключевая ставка (ставка рефинансирования) в РФ*

*Источник*: Банк России.

Оцененная модель производственной функции Кобба–Дугласа в целом сонаправлена с ВВП и позволяет достаточно точно предсказывать фазы роста и падения, однако неточно определяет их размер. Так, на протяжении первых 10 лет модель дает оценки выше реального ВВП, а на протяжении последних 10 лет — ниже реального выпуска. Тем не менее, для модели пересекающихся поколений, внимание которой акцентировано скорее на домохозяйствах, подобная оценка производственной функции будет приемлемой.

Оцененная в соответствии с уравнением (12) модель предсказания ставки процента достаточно точно показывает реальную ставку рефинансирования с 2000 по 2014 г., и лишь в последние годы оценка недостаточно точна (см. рис. 2). Для расчета ставки из уравнения (12) использовалась оценка коэффициента амортизации капитала $\delta = 0{,}05$, полученная в п. 3.2.

### 3.2. Подбор параметров $\{\delta,\theta,\sigma,\beta\}$

Последний неопределенный параметр задачи максимизации прибыли репрезентативных фирм (6), от которого зависит в том числе и ставка процента, — коэффициент амортизации капитала $\delta$. Следуя (Voskoboynikov, 2012), он был взят равным $\delta = 0{,}05$. В литературе приводится мало макрооценок этого параметра для экономики РФ. Данная оценка, несмотря на то что была сделана более 10 лет назад, — оказалась самой последней.

Эластичность предложения труда Фриша $\theta$ из (4), примем равной $1/1{,}77$, следуя работе (Shulgin, 2018), где она была оценена на микроданных; межвременная эластичность замещения потребления — $\sigma = 1{,}97$ (Shulgin, 2018; Charnavoki, 2019); субъективный коэффициент дисконтирования — $\beta = 0{,}905$ (Khvostova, Larin, Novak, 2014).

### 3.3. Теоретическое обоснование параметров $\{\tilde{l}, S, R, \tau, \chi_s^n\}$

Время, распределяемое каждый период каждым индивидом на труд и отдых, $\tilde{l} = l_{s,t} + n_{s,t}\ \forall s,t$, задано равным 1. Число когорт населения $S$ задано отталкиваясь от возрастных групп экономически активных индивидов, так как индивиды в нашей модели получают заработную плату, сберегают и т.д. Поэтому в рамках модели люди в возрасте до 20 лет не включаются в рассмотрение, так как их участие в экономике — около нуля, что следует из (Лайкам, 2022).

Таким образом, $S = 73 - 20 = 53$, где 73 — средняя продолжительность жизни в стране, а 20 — возраст, до которого граждане экономически неактивны.

Возраст выхода на пенсию будет взят усредненный по полу, так как в рамках модели нет разделения индивидов на мужчин и женщин. Начиная с 2028 г., когда кончится переходный период к новому пенсионному возрасту, пенсионные возраста составят 65 и 60 лет для мужчин и женщин соответственно. Тогда, $R = 0{,}5(60+65) - 20 = 42$, где 20 вычтено, чтобы привести пенсионный возраст к $\{s_i\}_{i=1}^{S=53}$ множеству когорт.

В модели не учитывается досрочный выход на пенсию, так как доля пенсионеров, получающих пенсию до достижения пенсионного возраста, мала и сокращается. В (Деркач, 2015) показано, что, начиная с 2010 и до 2014 г. доля досрочных пенсионеров сократилась с 15,4 до 13% общей численности пенсионеров, и эта тенденция продолжается. Так, в 2023 г. общая численность трудовых пенсионеров составила 41,7 млн человек, из которых только 4,5 млн человек не достигли общеустановленного пенсионного возраста, что составляет 10,7% пенсионеров[4]. Так как тенденция сокращения этой доли, вероятнее всего, продолжится и в будущем, то учитывать досрочных пенсионеров в рамках данной долгосрочной модели не представляется необходимым.

Ставка налогообложения, полностью расходуемого на выплаты пенсионерам подоходного трудового налога, $\tau^l = 0{,}22$, так как именно столько любой работодатель отчисляет в пенсионный фонд с заработной платы каждого работающего[5]. Ставки пенсионного налога на прибыль $\tau^c = 0$ и пенсионного налога на доход по процентам с вкладов $\tau^k = 0$, так как эти налоги не используются для выплат пенсий в начальном равновесии.

---

[4] СФР, Сведения о численности пенсионеров, состоящих на учете в системе СФР, получающих пенсии в соответствии с Федеральным законом «О страховых пенсиях», не достигших общеустановленного пенсионного возраста (в том числе работающих и неработающих).

[5] В соответствии со ст. 425 Налогового кодекса РФ.

В действительности порядка 2,8 трлн руб. из 8,8 трлн руб., расходуемых на выплату страховых пенсий по старости, Пенсионный фонд получает из источников, не связанных с соответствующим $\tau^l$ налогом. Межбюджетные трансферты ПФР на выплату страховых пенсий составляют порядка 1 трлн руб., остальные необходимые средства (1,8 трлн руб.) Пенсионный фонд получает из своих прочих доходов и нецелевых бюджетных трансфертов[6]. В целях моделирования данное софинансирование бюджета ПФР предполагается экзогенным и не учитывается в модели. Для упрощения модели и вычислений параметр $\chi_s^n$ ($s \in \{s_i\}_{i=1}^{S=53}$) принят за единицу.

**3.4.** Калибровка параметров $\{b, \upsilon\}$

Для оценки параметров эллиптической функции антиполезности труда из (2) $b$ и $\upsilon$ необходимо, чтобы она аппроксимировала функцию постоянной эластичности труда Фриша (4), где $\theta = 1/1{,}77 = 0{,}565$.

Оценки $b$ и $\upsilon$ получены с помощью обобщенного метода моментов, где в качестве момента была выбрана квадратичная норма предельной полезности труда:

$$L^2 = \left\| MU_{cfe} - MU_{ellip} \right\| = \Sigma \left( n^{1/\theta} - (b/\tilde{l})(n/\tilde{l})^{\upsilon-1} \left(1 - (n/\tilde{l})^{\upsilon}\right)^{(1-\upsilon)/\upsilon} \right)^2. \qquad (22)$$

Минимизировав (22) в границах $n_{s,t} \in \left[10^{-10}, 1+10^{-10}\right]$ за счет подбора параметров $b$ и $\upsilon$ методом L-BFGS (Limited-memory Broyden–Fletcher–Goldfarb–Shanno algorithm) (Zhu, Byrd, 1997), были получены оценки $b = 0{,}43$, $\upsilon = 1{,}765$.

Предельный эффект функции эллиптической антиполезности труда достаточно точно аппроксимирует предельный эффект функции полезности труда Фриша, что видно из графика на рис. 3.

Итоговая калибровка всех параметров представлена в табл. 1.

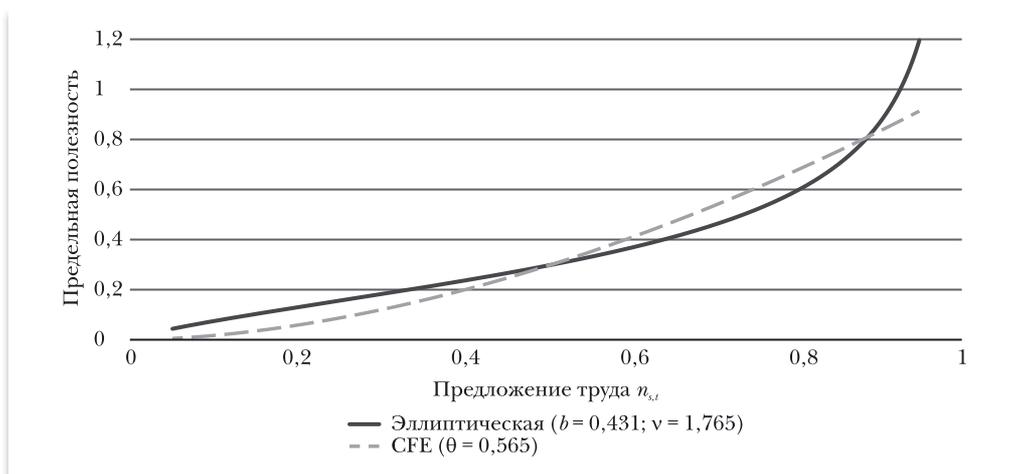

**Рис. 3.**
*Предельная полезность труда для функции Фриша и эллиптической функции*

---

[6] Структура доходов и расходов бюджета Пенсионного фонда Российской Федерации за 2022 г.

**Таблица 1.**

Калибровка параметров модели

| Параметр | Описание | Значение |
|---|---|---|
| α | Доля капитала в доходе | 0,3573 |
| A | Общая производительность факторов производства | 1,889 |
| δ | Попериодный коэффициент амортизации | 0,05 |
| θ | Коэффициент эластичности предложения труда Фриша | 0,565 |
| σ | Межвременная эластичность замещения | 1,97 |
| β | Субъективный коэффициент дисконтирования | 0,905 |
| $\tilde{l}$ | Запас времени в каждом периоде | 1 |
| s | Число периодов в жизни индивида | 53 |
| R | Период выхода на пенсию | 42 |
| $\tau^l$ | Ставка трудового налога в период $t = 0$ | 0,22 |
| $\tau^c$ | Ставка налога на прибыль в период $t = 0$ | 0 |
| $\tau^k$ | Ставка налога на проценты по вкладам в период $t = 0$ | 0 |
| $\chi^n_s$ | Коэффициент относительного масштаба полезности труда | 1 |
| b | Параметр эллиптической полезности труда | 0,431 |
| υ | Параметр эллиптической полезности труда | 1,765 |

### 4. Анализ результатов моделирования

В этом разделе откалиброванная (см. табл. 1) модель используется для симулирования экономических последствий разных сценариев перехода к накопительной пенсионной системе.

**4.1.** Базовая экономика

Прежде чем рассматривать влияние институциональных изменений и динамику перехода в новое равновесие, необходимо оценить качество подгонки модели базовой экономики (далее — базовой модели, т.е. модели для сравнения). Так как главная цель данной работы — анализ влияния отмены пенсионных выплат на благосостояние граждан, важно удостовериться в том, что модель точно описывает распределение предложения труда, от которого зависит доход, потребление и благосостояние граждан.

На рис. 4 показана структура рабочей силы по разным возрастным группам в данных и в базовой модели. Данные взяты из отчета Росстата о занятости в 2022 г. (Лайкам, 2022). В целом базовая модель демонстрирует схожую с данными динамику. Кривая данных и кривая модели выпуклы вверх по оси жизненного цикла, т.е. с возрастом индивиды начинают предлагать все меньше труда, — меньше всего работают люди пенсионного возраста. В модели молодые и старые индивиды предлагают несколько больше труда, чем в данных, — это отражено в предпосылках модели о том, что индивиды рождаются и входят в экономику без сбережений.

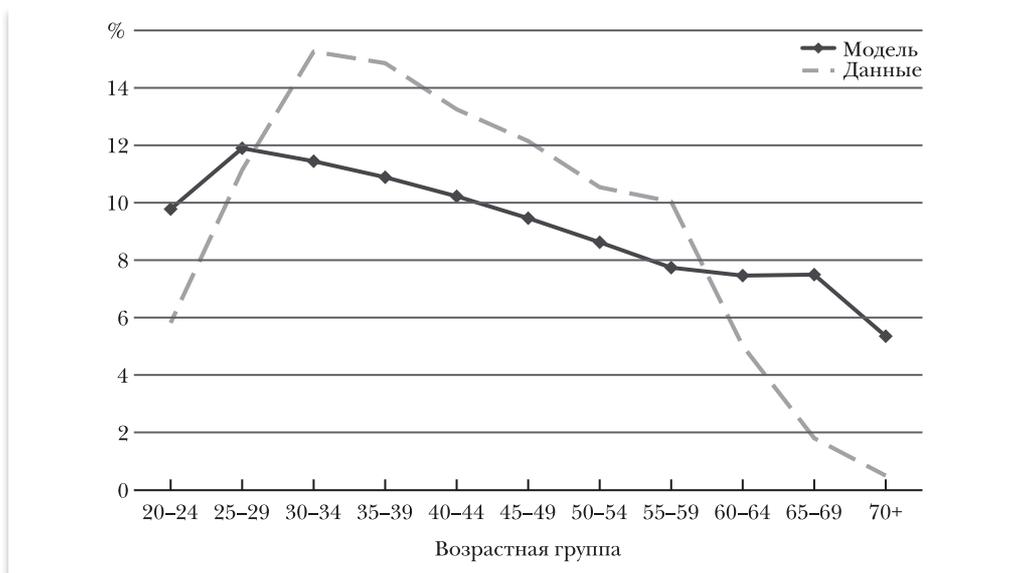

**Рис. 4.**
*Структура рабочей силы по возрастным группам*

Качество подгонки производственной функции можно оценить по рис. 1 и 2.

**4.2.** Проведение реформ и переходов в модели

Существует популярная концепция разделения пенсионных выплат на три типа, называемых «столпами» (Yermo, 2002). Первый «столп» представляет собой государственные фиксированные пенсионные выплаты, как правило, основанные на подоходном налоге, их целью является обеспечить в старости все, даже самые бедные слои населения, минимальным уровнем благосостояния. Второй «столп» — контролируемые самим работником схемы пенсионных выплат, предоставляющиеся в рамках трудового договора. В соответствии с концепцией Мирового банка второй «столп» также подразумевает создание накопительных счетов (Holzmann, 2005). Третий «столп» — частные сбережения на специальных пенсионных счетах, которые в том числе направлены на сглаживание во времени доходов обеспеченных слоев населения.

Так как второй «столп» почти не представлен в России, то в целях данной работы он будет концептуально объединен с третьим в рамках частных накоплений, управляемых гражданами, в том числе с закреплением в трудовом договоре.

Цель данной работы — *исследовать благосостояние во время перехода от солидарной пенсионной системы к накопительной*. Для того чтобы проанализировать все изменения ключевых показателей благосостояния, вводятся следующие сценарии реформ и для сравнения — базовая модель экономики.

1. «**Базовый сценарий**». Данный сценарий подразумевает отсутствие пенсионной реформы и сопутствующих изменений.

2. Сценарий перехода «**Трудовой налог**». В данной модели в 2024 г. отменяется солидарная пенсионная система и начинается 30-летний переходный период, в рамках которого всем вышедшим на пенсию до или во время этого периода пенсионерам будет выплачиваться минимальный размер оплаты труда, который с 2024 г. составляет 19 242 руб.[7] После окончания переходного периода выплаты продолжатся на уровне прожиточного минимума пенсионеров в размере 13 290 руб.[8] Выплаты МРОТ, а затем прожиточного минимума в данном случае представляют собой выплаты первого «столпа», в то время как платежи второго и третьего «столпов» представлены в виде частных накоплений индивидов. Финансирование этих выплат будет осуществляться при помощи трудового налога $\tau_{s,t}^l$ — как и прежде.

3. Сценарий перехода «**Налог на прибыль**». Дизайн модели аналогичен сценарию трудового налога, однако финансирование выплат будет осуществляться за счет налога на прибыль фирм $\tau_t^c$.

4. Сценарий перехода «**Налог с процентов по вкладам**». Дизайн модели аналогичен сценарию трудового налога. Однако финансирование выплат будет осуществляться за счет налога с процентов по вкладам индивидов $\tau_{s,t}^k$.

В сценариях 2–4 за размер выплат пенсионерам во время переходного периода берется МРОТ, а не прожиточный минимум, так как последний для пенсионеров значительно ниже, чем МРОТ. Его уровень часто критикуют за недостаточность для поддержания адекватной жизни[9]. Впоследствии выплаты первого «столпа» устанавливаются на уровне прожиточного минимума, так как подразумевается, что за время перехода индивиды успели накопить необходимые им средства для адекватных пенсий второго и третьего «столпов».

Также в литературе рассматриваются сценарии с финансированием перехода за счет внешнего долга (Makarski, Hagemejer, Tyrowicz, 2017), что могло бы стать удобным способом балансирования бюджета во время перехода, так как средняя ставка Мирового банка для кредитования России очень низкая и составляет $r_{wb} = 0{,}018$[10]. Моделирование подобного перехода показало, что за весь период соотношение внешнего долга к ВВП составило бы 600% и такой сценарий не является возможным, так как доступ к внешнему кредитованию в таких объемах (в том числе у Мирового банка) для России стал сильно затруднен после введения санкций в 2022 и 2023 г. Поэтому такой сценарий в дальнейшем не будет рассматриваться.

**4.3.** Макродинамика перехода

Проанализируем динамику переходного пути от солидарной пенсионной системы к накопительной в разных сценариях, все изменения показателей будут отображены на графиках. Отметим, что эти изменения рассчитываются путем сравнения значений в случае различных реформ со значениями, если реформа не проводилась (базовая модель). Будут обсуждаться изменения вплоть до 2124 г. (100 лет после 2024 г.), так как к тому моменту экономика приблизится к значениям в итоговом равновесии.

Для интерпретации результатов удобно будет поделить переходный период на три фазы.

---

[7] В соответствии с Федеральным законом от 27.11.2023 № 548-ФЗ.

[8] В соответствии с Федеральным законом от 27.11.2023 № 540-ФЗ.

[9] https://www.rbc.ru/economics/19/10/2020/5f8d79879a7947929631f66c

[10] World Bank (https://finances.worldbank.org/Loans-and-Credits/IBRD-Statement-of-Loans-Latest-Available-Snapshot/sfv5-tf7p).

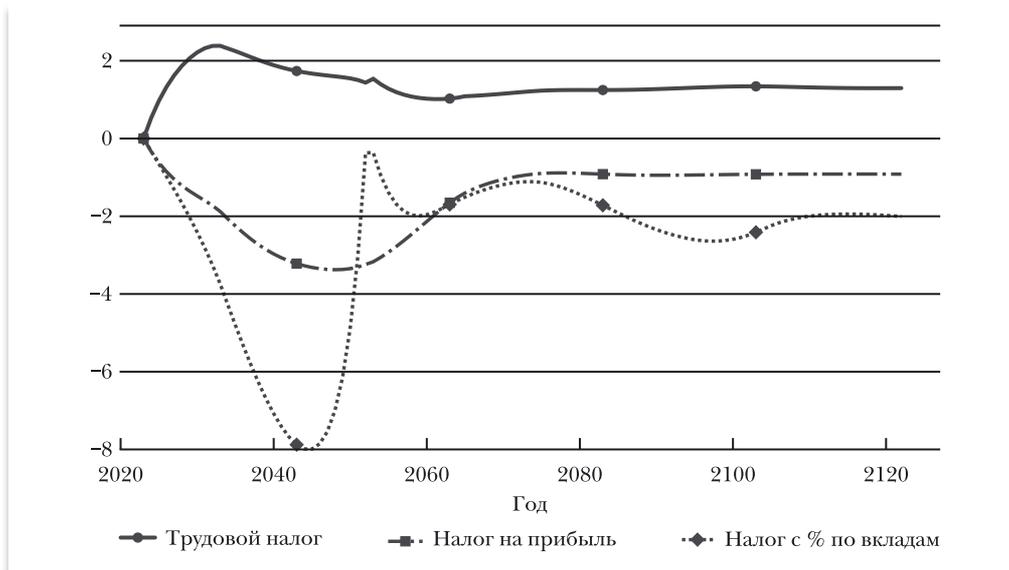

**Рис. 5.**
*Агрегированный выпуск, % изменений от уровня 2024 г.*

**Фаза 1** (2024–2054 гг.): переходный период, в его рамках все пенсионеры получают МРОТ.

**Фаза 2** (2054–2065 гг.): среди людей пенсионного возраста находятся как индивиды, получающие МРОТ, так и те, кто уже не получает государственных пенсионных выплат и живет на собственные сбережения.

**Фаза 3** (2065–2124 гг.): все пенсионеры в стране обеспечивают себя сами. Население страны все больше состоит из когорт, рожденных после внедрения реформы в 2024 г. И здесь достигается итоговое стационарное равновесие.

Для начала рассмотрим изменения в ВВП (рис. 5). Модель базовой экономики представляет собой горизонтальную линию в ноле. Видно, что сценарии с налогом на прибыль и с налогом на доход по вкладам в первой фазе сокращают ВВП, вероятно, из-за шока изменения налоговой пенсионной политики в стране, поскольку в то же время переход с налогов на прибыль в первой фазе заставляет производство расти в отсутствии налоговых изменений. Во второй фазе, после завершения переходного периода, сценарии с нетрудовым налогом демонстрируют рост ВВП. Этот рост замедляется в третьей фазе, однако нетрудовые сценарии так и не достигают дореформенного уровня. Сценарий с подоходным налогом в конечном итоге увеличивает ВВП примерно на 1,5%.

Следует отметить, что производство представляет собой композицию агрегированных капитала и предложения труда, которые сами зависят от других переменных. Чтобы понять механизмы внутри переходной динамики и определить их влияние на ВВП, рассмотрим динамику и других макроэкономических показателей.

На рис. 6 показаны изменения других агрегированных макроэкономических величин, капитала и предложения труда — слева и справа соответственно.

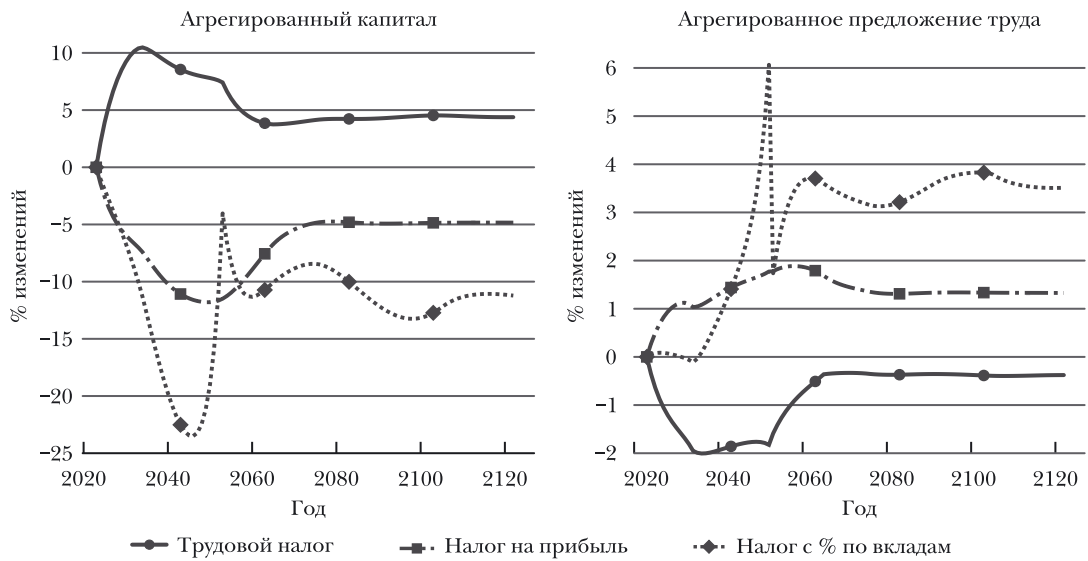

**Рис. 6.**
*Агрегированные макроэкономические показатели, % от базового уровня*

Динамика изменений агрегированного капитала в целом схожа с изменениями производства на рис. 5. Так как агрегированный капитал состоит из сбережений индивидов, мы видим также изменения накоплений населения во времени. В сценарии с использованием трудового налога население ожидаемо начинает больше сберегать, среднестатистические взрослые индивиды увеличивают сбережения, так как будут получать меньшую пенсию в будущем, молодые же наращивают накопления, так как в старости их пенсия будет еще меньше.

Сценарии с налогом на прибыль и налогом на процент по вкладам в первой фазе вызывают резкое сокращение капитала и сбережений. В случае налога на прибыль это происходит, из-за того что выплата процентов на капитал не подлежит вычету перед налогообложением и, таким образом, оптимальная величина капитала относительно числа занятых сокращается, поэтому фирмы снижают размеры заемных средств на пользу труда. В сценарии налога с процентов по вкладам индивидам в первой фазе становится значительно менее выгодно сберегать из-за появившегося налога на сбережения и накопления сокращаются. После окончания переходного периода налоги на прибыль и с процента по вкладам сокращаются, поэтому фирмам становится выгоднее арендовать капитал, а индивидам — привлекательнее открывать вклады. Здесь начинается стремительный рост сбережений. Этот рост во второй фазе затухает, по мере того как в экономике становится все больше индивидов, в среднем оптимально сберегавших всю жизнь, в третьем периоде все когорты становятся таковыми и достигается новое равновесие:

$$\frac{\partial Y_t}{\partial L_t} = A K_t^\alpha (1-\alpha) L_t^{-\alpha} \Rightarrow \frac{\partial^2 Y_t}{\partial L_t \partial K_t} = A \alpha K_t^{\alpha-1}(1-\alpha) L_t^{-\alpha} > 0. \tag{23}$$

На рис. 6 (справа) показано, как предложение труда в каждом сценарии изменяется обратным образом по отношению к капиталу, что можно объяснить предельной продуктивностью труда (23). В сценарии трудового налога производительность труда растёт, в прочих сценариях она сокращается. При финансировании перехода за счёт налога на прибыль фирмам становится выгоднее сместить производственный фокус на наём бо́льшего числа рабочих, поскольку выплаты заработных плат подлежат вычету, поэтому занятость растёт. Сценарий налога на процент по вкладам, как упоминалось ранее, сильно затрудняет процесс сбережения для населения, из-за этого работающие ориентируются больше на доход не от накоплений, а от заработных плат. Поэтому они вынужденно резко наращивают предложение труда, во второй фазе вместе с сокращением налога они стремительно увеличивают сбережения и сокращают свой труд.

В случае сценария перехода за счёт трудового налога, несмотря на сокращение ставки налога, предложение труда падает, что объясняется предельной продуктивностью труда (23). Также в каждом сценарии после начала перехода выплаты пенсионеров сокращаются относительно изначальных, из-за этого пожилые когорты населения наращивают предложение труда с целью возместить часть потерянного потребления. Однако в реальной жизни многие пенсионеры неспособны оптимально нарастить предложение труда из-за возраста, тогда благосостояние некоторых из них значительно ухудшится.

На рис. 7 показаны изменения в заработной плате и агрегированном потреблении. Тренд заработной платы следует за производством, так как в равновесии заработная плата равна предельной доходности труда $w_t = AK_t^{\alpha}(1-\alpha)L_t^{-\alpha}$, т.е. подобно выпуску полностью определяется уровнем сбережений и рабочей силы. Динамика также аналогична и развитию капитала (см. рис. 6). В равно-

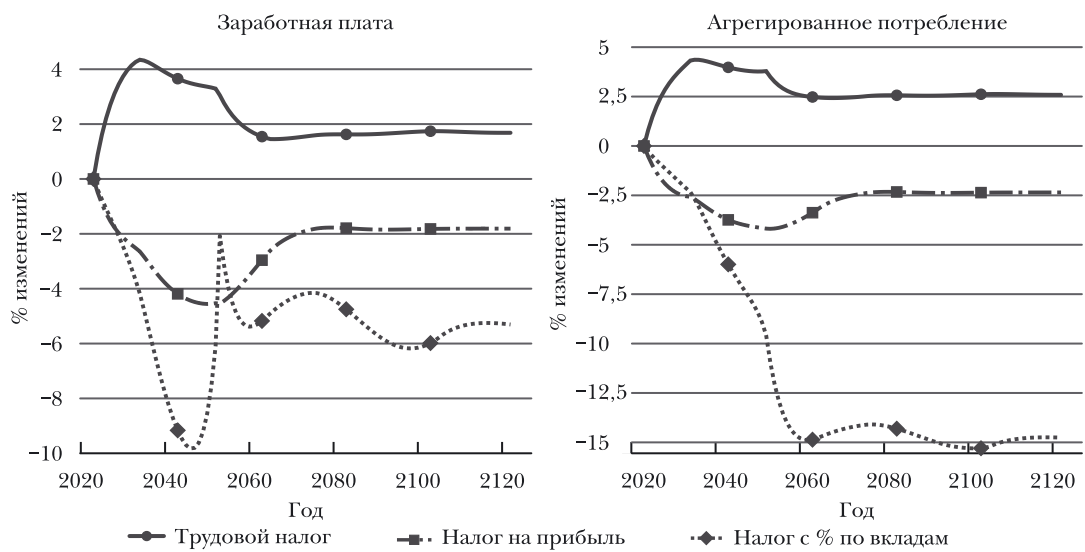

**Рис. 7.**
*Заработные платы и агрегированное потребление, % от базового уровня*

весии в сценарии финансирования перехода за счет трудового налога заработные платы растут. Однако в прочих сценариях заработные платы сокращаются, несмотря на рост занятости, что происходит из-за куда бо́льшего падения сбережений и капитала в той же фазе.

На правом графике на рис. 7 представлено агрегированное потребление. Сценарии с налогом на прибыль характеризуются умеренным падением потребления в первой фазе, умеренным ростом во второй и выходом в равновесие в третьей — на уровне ниже начального. Потребление же в случае налога на труд растет во всех фазах, кроме третьей. В сценарии налога на процент по вкладам потребление стремительно сокращается в первой фазе; во второй падение замедляется; в третьей потребление достигает нового равновесия, которое на 15% ниже начального. Для того чтобы увидеть влияние разных групп населения на итоговое потребление, необходимо рассмотреть потребление репрезентативных индивидов и возрастных групп, живших на момент реформы.

### 4.4. Анализ благосостояния

На рис. 8 приведено потребление за всю жизнь индивидов, живущих на момент реформы. Их возраст на момент реформы отложен на горизонтальной оси. Мы видим, что в сценариях без трудового налога молодые когорты в среднем проигрывают от реформы. Это происходит, из-за того что во время перехода сильно снижается реальная ставка по сбережениям (как следствие выросшего налога с процентов по вкладам в случае соответствующего сценария и как следствие этого — меньшей нормы дохода на капитал в производстве фирм в случае сценария налога на прибыль (см. рис. 6)). Молодые индивиды в первой половине жизни значительно меньше сберегают и впоследствии могут позволить себе меньше потребления. Для сценария с трудовым налогом канал шока

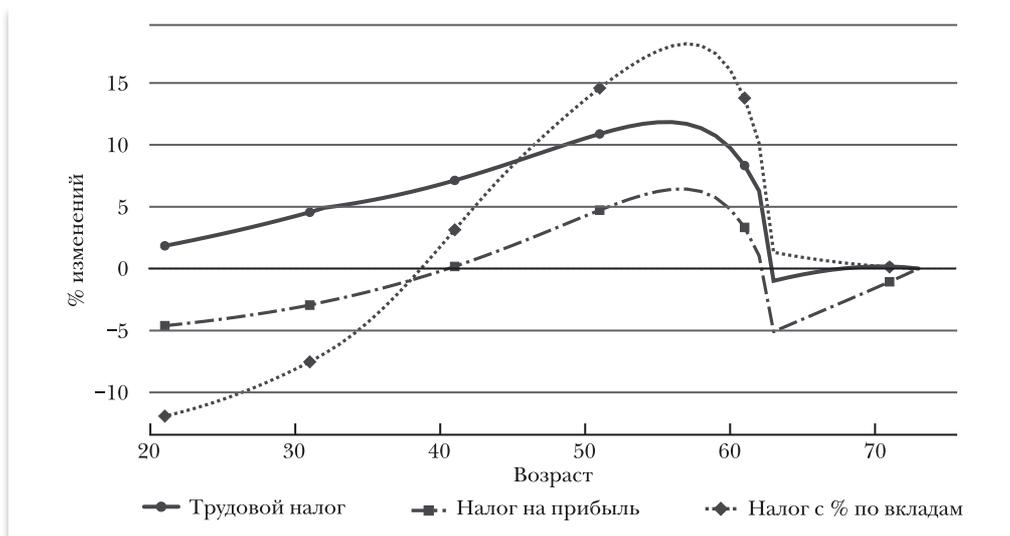

**Рис. 8.**
*Пожизненное потребление, % от базового уровня*

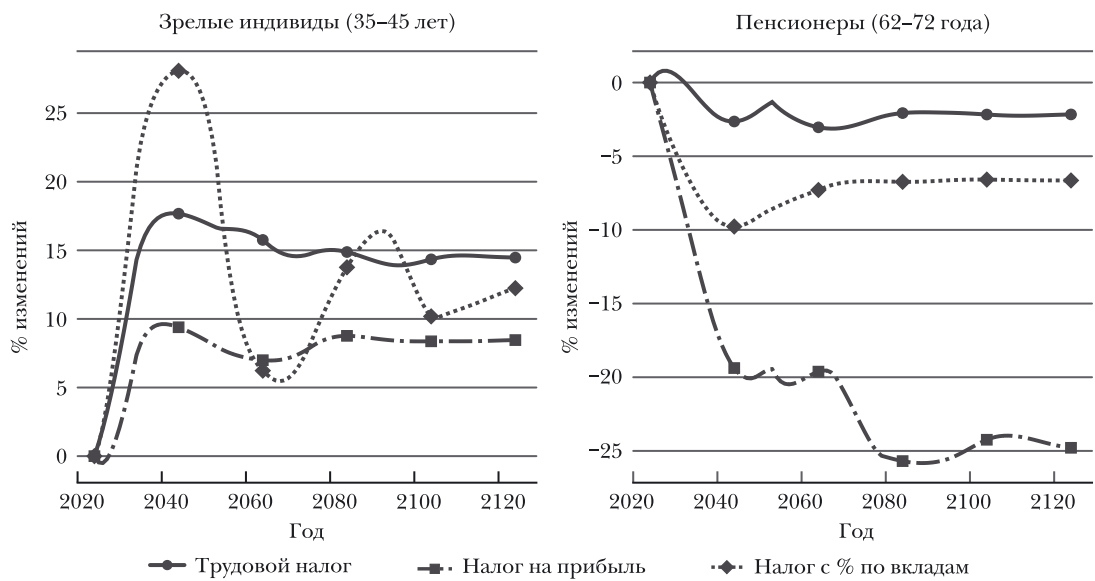

**Рис. 9.**
*Потребление репрезентативных индивидов, % от базового уровня*

доходности накоплений отсутствует и поэтому молодые индивиды закономерно улучшают жизненное потребление.

Средневозрастные и зрелые индивиды в среднем получают дополнительное пожизненное потребление во всех сценариях. В случае налога на труд это связано с тем фактом, что рабочие одновременно платят меньший налог, продолжают получать пенсии и сталкиваются с ростом производства (см. рис. 5), что увеличивает их доход. В прочих сценариях это объясняется относительной непривлекательностью инвестирования, вместо этого индивиды предпочитают заработанные деньги тут же конвертировать в потребление.

Среднестатистические пожилые индивиды получают незначительный рост или даже снижение потребления. Малый по модулю размер изменений объясняется тем, что они продолжают получать пенсию, пускай и уменьшенную, и никакие изменения в экономике не успевают сильно повлиять на их потребление в немногочисленные оставшиеся годы жизни.

Рис. 9 демонстрирует динамику усредненного потребления в рамках возрастной группы 35–45 лет и 62–72 года. Каждая реформа в среднем увеличивает потребление зрелых индивидов во всех фазах. Однако *все реформы уменьшают потребление пенсионеров.*

Зрелые индивиды после реформы сталкиваются с меньшим налогом и большим располагаемым доходом, что позволяет им в среднем больше потреблять. Притом эффект тем сильнее, чем меньше относительная ожидаемая привлекательность сбережения средств, поэтому в сценарии перехода с налогом на процент по вкладам наблюдается самый стремительный рост потребления и копить деньги становится относительно невыгодно. В первой фазе наблюда-

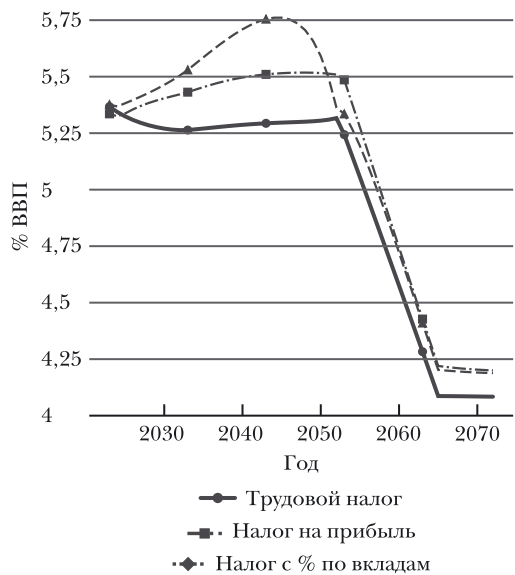

**Рис. 10.**
*Изымаемый налог по отношению к ВВП*

ется особенно большой рост потребления, так как в первые 30 лет после реформы индивиды могут рассчитывать на выплату им государственных пенсий на достаточно высоком уровне МРОТ и поэтому не стремятся значительно наращивать свои сбережения.

Для пенсионеров первая фаза перехода характеризуется самым большим падением потребления, что связано с сокращением государственных пенсионных выплат до уровня МРОТ и с тем фактом, что пенсионерами в первые десятки лет после реформы являются индивиды, сберегавшие неоптимально большую часть жизни. Во второй фазе потребление начинает в среднем медленно расти, несмотря на то что пенсионеров, получающих государственную пенсию на уровне МРОТ, все меньше, так как все большую часть вышедших на пенсию людей составляют индивиды, оптимально сберегавшие всю жизнь и скопившие достаточно средств на безбедную старость.

Сценарий трудового налога накладывает в среднем наименьшие бремя на пожилых индивидов, что достигается за счет отсутствия налогового искажения рынка сбережений, что позволяет индивидам во время переходного периода сберегать больше, чем в прочих сценариях, и войти в старость с более пригодным для безбедной жизни размером накоплений.

Сценарий налога на процент по вкладам характеризуется самым драматичным падением потребления пожилых индивидов. Это связано с тем, что во всех фазах и далее в новом равновесии существует налог на сбережения, делающий реальную доходность по ним недостаточной для целей большинства индивидов. Таким образом, пенсионные выплаты в рамках второго и третьего «столпов» значительно ниже, чем в прочих сценариях, и поэтому потребление пенсионеров в новом равновесии на четверть (см. рис. 9) ниже, чем было изначально.

Отметим, что, несмотря на средние показатели в реальности во всех сценариях, будут существовать *подгруппы пожилых индивидов за чертой бедности*, так как какая-то часть населения будет неспособна сберечь к старости необходимые средства. А пенсии первого «столпа» смогут обеспечить лишь минимальное потребление на *уровне выживания*.

Выводы о большей «мягкости» сценария перехода с применением трудового налога подтверждаются (рис. 10 демонстрирует величину изымаемого налога по отношению к ВВП). Нетрудно заметить, что трудовой налог на протяжении всего переходного периода накладывает наименьшее относительное налоговое бремя на фирмы и население, в том числе и на будущих пенсионеров.

Это происходит за счет того что в сценариях прочих налогов, в отличие от сценария трудового налога, первая фаза характеризуется спадом в агрегированном выпуске, в то время как при переходе с трудовым налогом ВВП растет, а значит, и доля налога относительно него сокращается.

### 5. Выводы

Солидарная пенсионная система является главным «столпом» пенсионной системы, в том числе и в России, она служит как страховка от рисков потерять сбережения перед старостью и как канал перераспределения богатств между поколениями. Но, несмотря на плюсы такой системы, ее становится невозможно поддерживать в странах с растущим числом пожилых людей и сокращением численности рабочей силы. Это относится к ситуации в России.

В данной работе представлен анализ того, как перейти от солидарной системы к более устойчивой системе пенсионных выплат — накопительной. В рамках исследования строится модель общего равновесия пересекающихся поколений для анализа динамики изменений показателей в переходном периоде в разных сценариях. Особое внимание уделяется рассмотрению эффектов, влияющих на благосостояние и макроэкономические показатели. Также детально показаны последствия такого перехода как для нынешних, так и для будущих поколений.

Проведенный анализ показал, что переход к накопительной пенсионной системе, финансируемый за счет налога на прибыль или налога на доход от сбережений, накладывает бремя на всех индивидов, живущих на момент реформы, они вынужденно сберегают меньше оптимального уровня, при этом начинают больше работать и меньше потреблять. В то же время переход с помощью трудового налога не сопряжен с такими тяжелыми последствиями и поэтому является, пожалуй, самым желательным.

Переход к накопительной системе особенно сильно ухудшает положение пенсионеров, живущих во время переходного периода. Однако в долгосрочной перспективе их положение почти не меняется при переходе к новой системе, финансируемой налогами на прибыль или на трудовые доходы.

Тем не менее, результаты моделирования справедливы для среднестатистических индивидов. В действительности благосостояние многих людей после реформы будет изменяться иначе. Многие люди в будущем не смогут скопить необходимые на старость средства и в пожилом возрасте столкнутся с бедностью.

Наконец показано, что плавный пенсионный переход почти не изменяет ВВП в долгосрочной перспективе. При переходе за счет трудового налога наблюдается экономический рост. Это происходит в первую очередь через канал сбережений, которые следуют вышеописанному тренду ВВП. В новом равновесии, т.е. после перехода к накопительной системе, заработная плата и занятость также незначительно возрастают в сценарии трудового налога, однако в прочих сценариях наблюдается их сокращение.

Стоит упомянуть, что моделирование осуществлялось на основе предпосылки о совершенном горизонте инвестиционного планирования индивидов. В реальности же данная предпосылка не соблюдается, работающие скорее проявляют «инвестиционную близорукость». Для решения этой проблемы можно ввести систему сбережения «по умолчанию», в рамках которой работодатель будет отчислять определенную долю заработной платы в выбранный индивидом пенсионный фонд. При этом у граждан будет возможность как отказаться от такой системы в пользу другой, так и изменить в ней ставку сберегаемой части заработной платы. Подобные пенсионные планы уже существуют во многих странах ОЭСР (Stéphanie, 2023), это позволяет значительно большей доле индивидов сберечь достаточную сумму к старости, несмотря на проблему «инвестиционной близорукости».

K.V. Moiseyev
HSE University, Nizhny Novgorod, Russia

# Modeling the transition from pay-as-you-go to a fully funded pension system in Russia[11]

**Abstract.** In countries with a growing number of elderly and a shrinking workforce, one of which is Russia, it becomes impossible to maintain a solidary pension system and a need to switch to a more stable funded system appears. This paper analyzes various scenarios of Russia's transition to such a system. This is the first study on the Russian economy in which an Overlapping Generations Model is used to simulate the pension transition. It is demonstrated that in the long term, the transition to a funded system slightly reduces the welfare of pensioners, and during the transition, the situation of pensioners deteriorates strongly. However, it is also important to emphasize that the transition imposes a heavy burden on all generations living during the reform, they are forced to consume less and greatly change their savings, while also often starting to work more. Such conclusions are made concerning average population cohorts, and the results may not be the same for different groups of individuals within these cohorts. In different scenarios, the pension system transition can cause both economic growth and economic recession, as well as a corresponding increase or decrease in wages and consumption.

**Keywords:** *pension reform, funded pension system, dynamic model, pay-as-you-go pension system, OLG model.*

JEL Classification: D58, E2, E6.

For reference: **Moiseyev K.V.** (2024). Modeling the transition from pay-as-you-go to a fully funded pension system in Russia. *Journal of the New Economic Association*, 3 (64), 30–52 (in Russian).

DOI: 10.31737/22212264_2024_3_30-52

EDN: NPVHJW

---